\documentclass{article}
\usepackage{amsmath}
\usepackage{amsfonts}
\usepackage{amssymb}
\usepackage{graphicx}

\input{tcilatex}
\begin{document}

\title{NSIs in Semileptonic~Rare Decays of Mesons Induced by Second
Generation of Quarks ($c$-quark and $s$-quark)}
\author{Shakeel Mahmood$^{(1)}$; Farida Tahir; Azeem Mir \\
%EndAName
\textit{Comsats Institute of Information Technology,}\\
\textit{\ Department of Physics, Park Road,\ Chek Shazad,Islamabad}\\
$^{({\footnotesize 1)}}${\footnotesize shakeel\_mahmood@hotmail.com}}
\date{}
\maketitle

\begin{abstract}
We study rare decays$~D_{s}^{+}\rightarrow D^{+}\upsilon \overline{\upsilon }%
,~B_{s}^{0}\rightarrow B^{0}\upsilon \overline{\upsilon }~$and $%
K^{+}\longrightarrow \pi ^{+}\overline{\upsilon }\upsilon ~$in the frame
work of NSIs. We calculate Branching ratios of these decays. We explore the
possibility for second generation of quarks in NSIs just$~$like leptonic
contribution in $\epsilon _{\alpha \beta }^{e},\epsilon _{\alpha \beta
}^{\mu }$and $\epsilon _{\alpha \beta }^{\tau }$. We study the dependence of 
$B_{s}^{0}\rightarrow B^{0}\upsilon \overline{\upsilon }$ on $\epsilon
_{\tau \tau }^{uL}$. We show that there exist a possibility for $\epsilon
_{\tau \tau }^{cL},$ also from these reactions. Three other processes $%
D^{+}\longrightarrow \pi ^{+}\overline{\upsilon }\upsilon
,D^{0}\longrightarrow \pi ^{0}\overline{\upsilon }\upsilon $ and $%
D_{s}^{+}\longrightarrow K^{+}\overline{\upsilon }\upsilon $ are
investigated with $s$ quark in the loop for NSIs instead of $d$ quark.
Constraints on $\epsilon _{\tau \tau }^{sL}~$and $\epsilon _{ll^{\prime
}}^{sL}~(l,l^{\prime }\neq \tau )~$are$~$provided. We point out that
constraints for both $u~$and $c$ quark are equal $(\epsilon _{\tau \tau
}^{uL}~=\epsilon _{\tau \tau }^{cL})~$and similarly for $d$ and $s$ quarks
the constraints are equal ($\epsilon _{\tau \tau }^{dL}~=\epsilon _{\tau
\tau }^{sL})$.

%TCIMACRO{\TeXButton{PACS}{PACS numbers:} }%
%BeginExpansion
PACS numbers:
%EndExpansion
12.60.-i, 13.15.+g, 13.20.-v
\end{abstract}

\section{Introduction}

Rare decays of mesons having two neutrinos in the final state are thought to
be a clean signal for the NP. These decays provide us a unique opportunity
to study NSIs. NSIs is thought to be a very well anticipated phenomena. The
effective Lagrangian for NSIs in model independent way is given in \cite{S.
Davison} and can be written as 
\begin{equation*}
L_{eff}^{NSI}=-2\sqrt{2}G_{F}\left[ \underset{\alpha =\beta }{\sum }\epsilon
_{\alpha \beta }^{fP}(\overline{\nu }_{\alpha }\gamma _{\mu }L\nu _{\beta })(%
\overline{f}\gamma ^{\mu }Pf)+\underset{\alpha \neq \beta }{\sum }\epsilon
_{\alpha \beta }^{fP}(\overline{\nu }_{\alpha }\gamma _{\mu }L\nu _{\beta })(%
\overline{f}\gamma ^{\mu }Pf)\right]
\end{equation*}%
Here $\epsilon _{\alpha \beta }^{fP}$ is the parameter for NSIs, which
carries information about dynamics. NSIs are thought to be well-matched with
the oscillation effects along with new features\ in neutrino searches \cite%
{P. Huber}\cite{P. Huber 1}\cite{P. Huber 2}\cite{P. Huber 3}\cite{P. Huber
4}\cite{P. Huber 5}\cite{P. Huber 6}. It is believed that NSIs can affect
neutrinos at production, propagation and detection level. Constraints on
NSIs parameter $\epsilon _{\alpha \beta }^{fP}$ have been studied in
References \cite{S. Davi}\cite{V. D. Barger}\cite{Z. Berezhian}. These are
loop induced interactions in standard model (SM), consisting of charge as
well as neutral vertices but NSIs will affect neutral vertices only \cite{A
Bandyopadhyay}. From scattering in leptonic sectors $(f$ is lepton$),$
constraints are determined for first two generations $\epsilon _{ll}^{fP}$ ($%
l=e$,$\mu $ ) by tree level processes and could be limited at $O(10^{-3})$
by future $\sin ^{2}\theta _{W}$ experiments. For third generation ($\tau $)
decays which occur at loop level are studied. The limit of $O(0.3)$ is
expected for the third generation ($\tau $) is KamLAND data \cite{Phys. Rev.
Lett} and solar neutrino data \cite{K. Eguchi et al.}\cite{J. Boger}.
Although, the constraints on $\epsilon _{\tau l}^{fP}$ are given by the
precision experiments but they are bounded by $O(10^{-2})$ \cite{M. L.
Mangano}. It is pointed out in reference \cite{C. H. Chen} that by using $%
K^{+}\longrightarrow \pi ^{+}\overline{\upsilon }\upsilon ~$the $\epsilon
_{\tau \tau }^{uL}$ constraints could be $O(10^{-2}).~$Mostly $f~~$is lepton
or quark from first generation ($u$ or $d$). If we take $f~$from second
generation of quark we have almost same constraints as for first
generations, $\epsilon _{\alpha \beta }^{cP}$ $\approx $ $\epsilon _{\alpha
\beta }^{uP}$. Similar thing happen to the other partner of $c$, $s$ quark
and we can have $\epsilon _{\alpha \beta }^{dP}$ $\approx $ $\epsilon
_{\alpha \beta }^{sP}.$ We inspire from leptonic sector where we have $%
\epsilon _{\alpha \beta }^{eP}$,$~\epsilon _{\alpha \beta }^{\mu P}\ $and
even $\epsilon _{\alpha \beta }^{\tau P}$. Although, no body is talking
about these types of effects for the second generation simply due to the
fact that the ordinary matter consist of only of first generation of quarks
but we point out that just like second generation of leptons NSIs are also
affected by second generation of quarks at the production of neutrinos from
rare decays of mesons. These could be responsible for the flavor violating
neutrino production.

We investigate $K^{+}\longrightarrow\pi^{+}\overline{\upsilon}\upsilon
,D_{s}^{+}\rightarrow D^{+}\overline{\upsilon}\upsilon~$and $%
B_{s}^{0}\rightarrow B^{0}\overline{\upsilon}\upsilon$ processes for the
this purpose. These processes$~$will be very important tool for the search
of possible new physics. Here we proceed as follows. We revise $%
K^{+}\longrightarrow\pi^{+}\overline{\upsilon}\upsilon$ in the SM as well in
NSIs for $c$ quark in the loop and then it is examined in NSIs with $c$ in
the loop. $B_{s}^{0}\rightarrow B^{0}\overline{\upsilon}\upsilon~$is studied
for the first time in NSIs with u quark. After this $D_{s}^{+}\rightarrow
D^{+}\overline{\upsilon}\upsilon~$and $B_{s}^{0}\rightarrow B^{0}\overline{%
\upsilon}\upsilon~$are searched for NSIs with $c$ quark. Just like these
three other processes $D^{+}\longrightarrow\pi^{+}\overline{\upsilon }%
\upsilon,D^{0}\longrightarrow\pi^{0}\overline{\upsilon}\upsilon$ and $%
D_{s}^{+}\longrightarrow K^{+}\overline{\upsilon}\upsilon$ calculated with $%
s $ quark instead of $d$ quark. Then results and comparison is provided and
conclusion is given at the end.

\section{Experimental Status}

It is expected that at the end of this decade we will be able to detect rare
decays of meson involving neutrinos in the final state just like $%
K^{+}\longrightarrow\pi^{+}\overline{\upsilon}\upsilon~$\cite{Andrzej J}.
But so far, it is the only semileptonic reaction involving two neutrinos in
the final state who experimental value is known which is $(1.7\pm1.1)\times
10^{-10}$\cite{J.Beringer et al}. So by using this reaction we can point out
exact region for the new physics. $D_{s}^{+}\rightarrow D^{+}\overline {%
\upsilon}\upsilon,$ $B_{s}^{0}\rightarrow B^{0}\overline{\upsilon}%
\upsilon,D^{+}\longrightarrow\pi^{+}\overline{\upsilon}\upsilon,D^{0}%
\longrightarrow\pi^{0}\overline{\upsilon}\upsilon$ and $D_{s}^{+}%
\longrightarrow K^{+}\overline{\upsilon}\upsilon$ are yet to be detected. In
super b-factories and in future super collider, we will have an opportunity
to detect them in a clean environment.

\section{Standard Model Calculations}

These reactions are represented by the quark level process $\overline {s}%
\longrightarrow\overline{d}$ $\overline{\upsilon}\upsilon$ for which penguin
diagram is

\begin{tabular}{l}
$\underset{\text{%
%TCIMACRO{\TeXButton{Fig}{\ Figure}}%
%BeginExpansion
\ Figure%
%EndExpansion
}}{\FRAME{itbpF}{1.5783in}{1.2004in}{0in}{}{}{Figure}{\special{ language
"Scientific Word"; type "GRAPHIC"; maintain-aspect-ratio TRUE; display
"USEDEF"; valid_file "T"; width 1.5783in; height 1.2004in; depth 0in;
original-width 1.542in; original-height 1.1666in; cropleft "0"; croptop "1";
cropright "1"; cropbottom "0"; tempfilename
'NFE53W01.wmf';tempfile-properties "XPR";}} }$%
\end{tabular}

%TCIMACRO{\TeXButton{Comment}{\ {}{}{}}}%
%BeginExpansion
\ {}{}{}%
%EndExpansion

along with two box diagrams not shown here, and the effective hamiltonian is

\begin{equation*}
H_{eff}^{SM}=\frac{G_{F}}{\sqrt{2}}\frac{\alpha_{em}}{2\pi\sin^{2}\theta_{W}}%
\underset{\alpha,\beta=e,\mu,\tau}{\Sigma}(V_{cd}^{%
\ast}V_{cs}X_{NL}^{l}+V_{td}^{\ast}V_{ts}X(x_{t}))\times(\overline{s}%
d)_{V-A}(\nu_{\alpha }\overline{\nu}_{\beta})_{V-A}
\end{equation*}
where $X_{NL}^{l}$ is the charm quark contribution. $X(x_{t})~$is
representing top quark contribution. Such processes are dominated by short
distance because long distance contribution are almost $10^{-3}$ less than
short distance. The up type quark in loop will increase the Branching ratios
of these reactions. The causes of uncertainties for these types of reactions
are $CKM$ matrix elements and hadronic uncertainties but for $%
K^{+}\longrightarrow\pi ^{+}\overline{\upsilon}\upsilon,D_{s}^{+}\rightarrow
D^{+}\overline{\upsilon }\upsilon~$and $B_{s}^{0}\rightarrow B^{0}\overline{%
\upsilon}\upsilon$ \ hadronic uncertainties can be eliminated by normalizing
with tree level processes. So, these are theoretically clean processes in SM
and due to loop they are very attractive for new physics.

SM Br of $K^{+}\longrightarrow\pi^{+}\overline{\upsilon}\upsilon$ is given by

$\frac{Br(K^{+}\rightarrow\pi^{+}\overline{\upsilon}\upsilon)}{%
Br(K^{+}\rightarrow\pi^{0}e^{+}\upsilon)}=r_{K^{+}}\frac{\alpha_{em}^{2}}{%
|V_{us}|^{2}2\pi^{2}\sin^{4}\theta_{W}}\underset{\alpha,\beta=e,\mu,\tau}{%
\Sigma }|V_{cd}^{\ast}V_{cs}X_{NL}^{l}+V_{td}^{\ast}V_{ts}X(x_{t})|^{2}$

We get the branching ratios (Br) for such reactions by normalizing with a
tree level process which is linked with $K^{+}\longrightarrow\pi^{+}%
\overline {\upsilon}\upsilon$ by isospin symmetry.

As $\langle\pi^{+}\left\vert (\overline{s}d)_{V-A}\right\vert K^{+}\rangle=%
\sqrt{2}\langle\pi^{0}\left\vert (\overline{s}u)_{V-A}\right\vert
K^{+}\rangle$

with $r_{K^{+}}=0.901$ is isospin effect given in \cite{W. Marciano} ;

Using $V_{us}=0.2252$; $V_{ud}=0.97425;$ $\theta_{w}$ $=28.7^{%
\circ};BR(K^{+}\longrightarrow\pi^{0}e^{+}\nu_{e})=5.07\times10^{-2}$\cite%
{PDG} the SM Br of $K^{+}\longrightarrow\pi^{+}\overline{\upsilon}\upsilon~$%
becomes $(7.8\pm0.8)\times10^{-11}$\cite{J. Brod et al.}

The margin for NSIs in $K^{+}\longrightarrow\pi^{+}\overline{\upsilon}%
\upsilon~$is equal to the difference of Br of theory and experiments which
is equal to $(0.92\pm1.18)\times10^{-10}$ approximately equal $10^{-10}.~$SM
Br for $D_{s}^{+}\rightarrow D^{+}\overline{\upsilon}\upsilon~$and $%
B_{s}^{0}\rightarrow B^{0}\overline{\upsilon}\upsilon$ is calculated as

$\frac{Br(D_{s}^{+}\rightarrow D^{+}\overline{\upsilon}\upsilon)}{%
Br(D_{s}^{+}\rightarrow D^{0}e^{+}\upsilon)}=\frac{\alpha_{em}^{2}}{%
|V_{us}|^{2}2\pi^{2}\sin^{4}\theta_{W}}\underset{\alpha,\beta=e,\mu,\tau}{%
\Sigma}|V_{cd}^{\ast}V_{cs}X_{NL}^{l}+V_{td}^{\ast}V_{ts}X(x_{t})|^{2}$

and

$\frac{Br(B_{s}^{0}\rightarrow B^{0}\overline{\upsilon}\upsilon)}{%
Br(B_{s}^{0}\rightarrow B^{+}e^{+}\upsilon)}=\frac{\alpha_{em}^{2}}{%
|V_{us}|^{2}2\pi^{2}\sin^{4}\theta_{W}}\underset{\alpha,\beta=e,\mu,\tau}{%
\Sigma}|V_{cd}^{\ast}V_{cs}X_{NL}^{l}+V_{td}^{\ast}V_{ts}X(x_{t})|^{2}$

Although for $Br(D_{s}^{+}\rightarrow D^{0}e^{+}\upsilon)$ and $%
Br(B_{s}^{0}\rightarrow B^{+}e^{-}\upsilon)~$we do not have experimentally
calculated values just like $~Br(K^{+}\rightarrow\pi^{0}e^{+}\upsilon)~$but
we have very elegantly estimated values for BES-III given in \cite%
{Eur.Phys.J} and we use them in our calculations. Here we are ignoring
effects of isospin breaking $D$ and $B~$mesons.

Using $Br(D_{s}^{+}\rightarrow
D^{0}e^{+}\upsilon)=5\times10^{-6},Br(B_{s}^{0}\rightarrow
B^{+}e^{-}\upsilon)=4.46\times10^{-8}$we get

$Br(D_{s}^{+}\rightarrow D^{+}\overline{\upsilon}\upsilon)_{SM}=7.72\times
10^{-15}$

$Br(B_{s}^{0}\rightarrow B^{0}\overline{\upsilon}\upsilon)_{SM}=6.86\times
10^{-17}$

\section{NSIs with u quark in the loop}

The NSIs effective hamiltonian is given by 
\begin{equation*}
H_{eff}^{NSI}=\frac{G_{F}}{\sqrt{2}}(V_{us}^{\ast}V_{ud}\frac{\alpha_{em}}{%
4\pi\sin^{2}\theta_{W}}\epsilon_{\alpha\beta}^{uL}\ln\frac{\Lambda}{m_{w}}%
)\times(\nu_{\alpha}\overline{\nu}_{\beta})_{V-A}(\overline{s}d)
\end{equation*}
from which the NSIs Br

$Br$($K^{+}\longrightarrow\pi^{+}\overline{\upsilon}\upsilon$)$%
_{NSI}=r_{K^{+}}\frac{\alpha_{em}^{2}}{|V_{us}|^{2}2\pi^{2}\sin^{4}\theta_{W}%
}|V_{us}^{\ast}V_{ud}\frac{1}{2}\epsilon_{\alpha\beta}^{uL}\ln\frac{\Lambda 
}{m_{w}}|^{2}\times$

$\ \ \ \ \ \ \ \ \ \ \ \ \ \ \ \ \ \ \ \ \ \ \ \ \ \ \ \ \ \ \ \ \ \ \
BR(K^{+}\longrightarrow\pi^{0}e^{+}\nu_{e})$

This was calculated in \cite{C. H. Chen} and the writers claimed that $%
\epsilon_{\tau\tau}^{uL}~\leq\frac{8.8\times10^{-3}}{\ln\frac{\Lambda}{m_{W}}%
}~.$With latest values $\epsilon_{\tau\tau}^{uL}~\leq\frac{6.7\times10^{-3}}{%
\ln\frac{\Lambda}{m_{W}}}.~$When we insert this value for our processes, we
have

$Br(B_{s}^{0}\rightarrow B^{0}\overline{\upsilon}\upsilon)_{NSI}=\frac {%
\alpha_{em}^{2}}{|V_{us}|^{2}2\pi^{2}\sin^{4}\theta_{W}}|V_{us}^{\ast}V_{ud}%
\frac{1}{2}\epsilon_{\alpha\beta}^{uL}\ln\frac{\Lambda}{m_{w}}|^{2}\times$

$\ \ \ \ \ \ \ \ \ \ \ \ \ \ \ \ \ \ \ \ \ \ \ \ \ \ \ \ \ \ \ \ \ \
Br(B_{s}^{0}\rightarrow B^{+}e^{-}\upsilon)$

Numerically we get

$Br(B_{s}^{0}\rightarrow B^{0}\overline{\upsilon}\upsilon)_{NSI}=2.17%
\times10^{-17}$

The $Br(D_{s}^{+}\rightarrow D^{+}\overline{\upsilon}\upsilon)_{NSI}=2.70%
\times10^{-15}$ is given in \cite{shakeel}

\section{NSIs with c quark in the loop}

Now we take $c$ quark in the loop instead of $u$ quark

\begin{tabular}{l}
$\underset{\text{%
%TCIMACRO{\TeXButton{Fig 1}{\ Fig 1.}}%
%BeginExpansion
\ Fig 1.%
%EndExpansion
}}{\FRAME{itbpF}{1.5039in}{1.2004in}{0in}{}{}{Figure}{\special{ language
"Scientific Word"; type "GRAPHIC"; maintain-aspect-ratio TRUE; display
"USEDEF"; valid_file "T"; width 1.5039in; height 1.2004in; depth 0in;
original-width 1.4684in; original-height 1.1666in; cropleft "0"; croptop
"1"; cropright "1"; cropbottom "0"; tempfilename
'NFE53W02.wmf';tempfile-properties "XPR";}} }$%
\end{tabular}

The NSIs effective hamiltonian is given by 
\begin{equation*}
H_{eff}^{NSI}=\frac{G_{F}}{\sqrt{2}}(V_{cs}^{\ast }V_{cd}\frac{\alpha _{em}}{%
4\pi \sin ^{2}\theta _{W}}\epsilon _{\alpha \beta }^{cL}\ln \frac{\Lambda }{%
m_{w}})\times (\nu _{\alpha }\overline{\nu }_{\beta })_{V-A}(\overline{s}d)
\end{equation*}%
Here, it is same as that of $u$ quark in the loop and we are simply
replacing $c$ with $u$. The NSIs Br with $c$ quark becomes

$Br$($K^{+}\longrightarrow\pi^{+}\overline{\upsilon}\upsilon$)$%
_{NSI}=r_{K^{+}}\frac{\alpha_{em}^{2}}{|V_{us}|^{2}2\pi^{2}\sin^{4}\theta_{W}%
}|V_{cs}^{\ast}V_{cd}\frac{1}{2}\epsilon_{\alpha\beta}^{cL}\ln\frac{\Lambda 
}{m_{w}}|^{2}\times$

$\ \ \ \ \ \ \ \ \ \ \ \ \ \ \ \ \ \ \ \ \ \ \ \ \ \ \ \ \ \ \ \ \ \ \
BR(K^{+}\longrightarrow\pi^{0}e^{+}\nu_{e})$

Putting the values from \cite{PDG} get calculate $\epsilon _{\tau \tau
}^{cL}~\leq \frac{6.2\times 10^{-3}}{\ln \frac{\Lambda }{m_{W}}}~.$When we
insert this value for our processes, we have

$Br(D_{s}^{+}\rightarrow D^{+}\overline{\upsilon}\upsilon)_{NSI}=\frac {%
\alpha_{em}^{2}}{|V_{us}|^{2}2\pi^{2}\sin^{4}\theta_{W}}|V_{cs}^{\ast}V_{cd}%
\frac{1}{2}\epsilon_{\alpha\beta}^{cL}\ln\frac{\Lambda}{m_{w}}|^{2}\times$

$\ \ \ \ \ \ \ \ \ \ \ \ \ \ \ \ \ \ \ \ \ \ \ \ \ \ \ \ \ \ \ \ \ \ \ \
Br(D_{s}^{+}\rightarrow D^{0}e^{+}\upsilon)$

$Br(B_{s}^{0}\rightarrow B^{0}\overline{\upsilon}\upsilon)_{NSI}=\frac {%
\alpha_{em}^{2}}{|V_{us}|^{2}2\pi^{2}\sin^{4}\theta_{W}}|V_{cs}^{\ast}V_{cd}%
\frac{1}{2}\epsilon_{\alpha\beta}^{cL}\ln\frac{\Lambda}{m_{w}}|^{2}\times$

$\ \ \ \ \ \ \ \ \ \ \ \ \ \ \ \ \ \ \ \ \ \ \ \ \ \ \ \ \ \ \ \ \ \
Br(B_{s}^{0}\rightarrow B^{+}e^{-}\upsilon)$

$Br(D_{s}^{+}\rightarrow D^{+}\overline{\upsilon}\upsilon)_{NSI}=2.57%
\times10^{-15}$

$Br(B_{s}^{0}\rightarrow B^{0}\overline{\upsilon}\upsilon)_{NSI}=2.0%
\times10^{-17}$

\section{ Rare Decays of D in The Standard Model}

SM Hamiltonian for short distance contribution of $c\rightarrow u\nu 
\overline{\nu}$ is given by 
\begin{equation*}
H_{eff}^{SM}=\frac{G_{F}}{\sqrt{2}}\frac{\alpha_{em}}{2\pi\sin^{2}\theta_{W}}%
\underset{\alpha,\beta=e,\mu,\tau}{\Sigma}[V_{cs}^{%
\ast}V_{us}X(x_{s})+V_{cb}^{\ast}V_{ub}X(x_{b})]\times(\overline{u}%
c)_{V-A}(\nu_{\alpha }\overline{\nu}_{\beta})_{V-A}
\end{equation*}
but for $D_{s}^{+}\rightarrow K^{+}\upsilon_{\alpha}\overline{\upsilon}%
_{\beta},~D^{+}\longrightarrow\pi^{+}\nu_{\alpha}\overline{\nu}_{\beta}$ and$%
~D^{0}\longrightarrow\pi^{0}\nu_{\alpha}\overline{\nu}_{\beta}$ the dominant
contribution comes from long distance. It is free from QCD complications
because they can be normalized with tree level process$.~$Their SM
contribution is given in table 2.

\section{NSIs in $D_{s}^{+}\rightarrow K^{+}\protect\upsilon_{\protect\alpha}%
\overline {\protect\upsilon}_{\protect\beta},~D^{+}\longrightarrow\protect\pi%
^{+}\protect\nu_{\protect\alpha}\overline{\protect\nu }_{\protect\beta}$ and$%
~D^{0}\longrightarrow\protect\pi^{0}\protect\nu_{\protect\alpha}\overline{%
\protect\nu}_{\protect\beta}$}

The quark level process $c\longrightarrow u\nu_{\alpha}\overline{\nu}%
_{\beta} $ is representing all above processes. For $D^{+}\longrightarrow%
\pi^{+}\nu_{\alpha}\overline{\nu}_{\beta},$ NSIs with $u$\ quark in the loop
it is calculated in \cite{C. H. Chen}

\begin{equation*}
Br(D^{+}\longrightarrow\pi^{+}\nu_{\alpha}\overline{\nu}_{%
\beta})_{NSI}=|V_{ud}^{\ast}\frac{\alpha_{em}}{4\pi\sin^{2}\theta_{W}}%
\epsilon_{\alpha\beta}^{dL}\ln\frac{\Lambda}{m_{W}}|^{2}BR(D^{+}%
\longrightarrow\pi^{0}e^{+}\nu_{e})
\end{equation*}
$Br(D^{+}\longrightarrow\pi^{+}\nu_{\alpha}\overline{\nu}_{%
\beta})_{NSI}=3.20\times10^{-8}|\epsilon_{\alpha\beta}^{dL}\ln\frac{\Lambda}{%
m_{W}}|^{2}$and it is mentioned that as $\alpha$ and $\beta$ could represent
any lepton, we take $\epsilon_{\tau\tau}^{dL}\sim1,\epsilon_{ll^{/}}^{dL}%
\langle1~$for $l=l^{/}\neq\tau.$ Here $\ln\frac{\Lambda}{m_{W}}\sim1.$

If we take $s$ quark in the loop instead of $d$ quark, then it is given by

\begin{tabular}{l}
$\underset{\text{%
%TCIMACRO{\TeXButton{Fig 2}{\  Figure  2.}}%
%BeginExpansion
\  Figure  2.%
%EndExpansion
}~\ }{\FRAME{itbpF}{1.5152in}{1.2955in}{0in}{}{}{Figure}{\special{language
"Scientific Word";type "GRAPHIC";maintain-aspect-ratio TRUE;display
"USEDEF";valid_file "T";width 1.5152in;height 1.2955in;depth
0in;original-width 1.4788in;original-height 1.26in;cropleft "0";croptop
"1";cropright "1";cropbottom "0";tempfilename
'NFE53W03.wmf';tempfile-properties "XPR";}}}$%
\end{tabular}

%TCIMACRO{\TeXButton{Comment}{\ {}{}{}}}%
%BeginExpansion
\ {}{}{}%
%EndExpansion
\begin{equation*}
H_{c\longrightarrow u\nu_{\alpha}\overline{\nu}_{\beta}}^{NSI}=\frac{G_{F}}{%
\sqrt{2}}(\frac{\alpha_{em}}{4\pi\sin^{2}\theta_{W}}V_{cs}V_{us}^{\ast
}\epsilon_{\alpha\beta}^{sL}\ln\frac{\Lambda}{m_{W}})(\overline{\nu}_{\alpha
}\nu_{\beta})_{V-A}(\overline{c}u)_{V-A}
\end{equation*}
and $Br$ becomes 
\begin{equation*}
Br(D^{+}\longrightarrow\pi^{+}\nu_{\alpha}\overline{\nu}_{\beta})_{NSI}=|%
\frac{V_{us}^{\ast}V_{cs}}{V_{cd}}\times\frac{\alpha_{em}}{4\pi\sin
^{2}\theta_{W}}\epsilon_{\alpha\beta}^{sL}\ln\frac{\Lambda}{m_{W}}%
|^{2}BR(D^{+}\longrightarrow\pi^{0}e^{+}\nu_{e})
\end{equation*}
$Br(D^{+}\longrightarrow\pi^{+}\nu_{\alpha}\overline{\nu}_{%
\beta})_{NSI}=3.25\times10^{-8}|\epsilon_{\alpha\beta}^{sL}\ln\frac{\Lambda}{%
m_{W}}|^{2}$and it is mentioned that as $\alpha$ and $\beta$ could represent
any lepton, we take $\epsilon_{\tau\tau}^{sL}\sim1,\epsilon_{ll^{/}}^{sL}%
\langle1~$for $l=l^{/}\neq\tau.$ Here $\ln\frac{\Lambda}{m_{W}}\sim1.$We
further see that same is applicable to two other processes $%
D_{s}^{+}\longrightarrow K^{+}\nu_{\alpha}\overline{\nu}_{\beta}~$and $%
D^{0}\rightarrow\pi^{0}\nu_{\alpha}\overline{\nu}_{\beta}.$

\begin{align*}
Br(D_{s}^{+} & \longrightarrow K^{+}\nu_{\alpha}\overline{\nu}_{\beta
})_{NSI}=|\frac{V_{us}^{\ast}V_{cs}}{V_{cd}}\frac{\alpha_{em}}{4\pi\sin
^{2}\theta_{W}}\epsilon_{\alpha\beta}^{sL}\ln\frac{\Lambda}{m_{W}}%
|^{2}BR(D_{s}^{+}\longrightarrow K^{0}e^{+}\nu_{e}) \\
Br(D^{0} & \longrightarrow\pi^{0}\nu_{\alpha}\overline{\nu}_{\beta})_{NSI}=|%
\frac{V_{us}^{\ast}V_{cs}}{V_{cd}}\frac{\alpha_{em}}{4\pi\sin ^{2}\theta_{W}}%
\epsilon_{\alpha\beta}^{sL}\ln\frac{\Lambda}{m_{W}}|^{2}BR(\overline{D}%
^{0}\longrightarrow\pi^{-}e^{+}\nu_{e})
\end{align*}
Using PDG 2012 \cite{PDG} Values $BR(D_{s}^{+}\longrightarrow
K^{0}e^{+}\nu_{e})=(3.7\pm1)\times10^{-3},$ $V_{ud}=0.97425\pm0.00022,$ $%
\alpha _{em}=\frac{1}{137},$we get

\begin{equation*}
Br(D_{s}^{+}\longrightarrow K^{+}\nu_{\alpha}\overline{\nu}%
_{\beta})_{NSI}=2.28\times10^{-8}(\epsilon_{\alpha\beta}^{sL})^{2}|\ln\frac{%
\Lambda }{m_{W}}|^{2}
\end{equation*}
For $\epsilon_{\tau\tau}^{sL}\sim1~$and $\ln\frac{\Lambda}{m_{W}}\sim1~,$we
get $BR(D_{s}^{+}\longrightarrow K^{+}\nu_{\alpha}\overline{\nu}_{\beta
})_{NSI}=0.14\times10^{-8}.$

Similarly for$~Br(\overline{D}^{0}\longrightarrow \pi ^{-}e^{+}\nu
_{e})=2.89\times 10^{-3}$we have 
\begin{equation*}
Br(D^{0}\longrightarrow \pi ^{0}\nu _{\alpha }\overline{\nu }_{\beta
})_{NSI}=3.25\times 10^{-8}(\epsilon _{\alpha \beta }^{sL})^{2}|\ln \frac{%
\Lambda }{m_{W}}|^{2}
\end{equation*}%
$10^{-8}$will be the reach of BES-III, so it is hopped that we might observe
these decays there. If not, even than useful limits for new physics can be
suggested. NSIs with $d$ quark are discussed for $D_{s}^{+}\longrightarrow
K^{+}\nu _{\alpha }\overline{\nu }_{\beta },$ and$~D^{0}\longrightarrow \pi
^{0}\nu _{\alpha }\overline{\nu }_{\beta }~$in \cite{shakeel}. Both values
are summarized in table two for comparison.

$\ \ \ \ \ \ \ $

\section{Results and Summary}

\begin{tabular}{|l|}
\hline
$\underset{\text{%
%TCIMACRO{\TeXButton{Plot1}{\ NSIs with u Plot}}%
%BeginExpansion
\ NSIs with u Plot%
%EndExpansion
}}{\FRAME{itbpF}{3.506in}{2.437in}{0in}{}{}{Figure}{\special{language
"Scientific Word";type "GRAPHIC";display "USEDEF";valid_file "T";width
3.506in;height 2.437in;depth 0in;original-width 3.7498in;original-height
2.3955in;cropleft "0";croptop "1";cropright "1";cropbottom "0";tempfilename
'NFE53W04.wmf';tempfile-properties "XPR";}}}$ \\ \hline
\end{tabular}

\begin{tabular}{|l|}
\hline
$\underset{\text{%
%TCIMACRO{\TeXButton{Plot}{\ NSIs with c}}%
%BeginExpansion
\ NSIs with c%
%EndExpansion
}}{\FRAME{itbpF}{3.4964in}{2.6775in}{0in}{}{}{Figure}{\special{language
"Scientific Word";type "GRAPHIC";display "USEDEF";valid_file "T";width
3.4964in;height 2.6775in;depth 0in;original-width 3.9375in;original-height
2.6351in;cropleft "0";croptop "1";cropright "1";cropbottom "0";tempfilename
'NFE53W05.wmf';tempfile-properties "XPR";}}}$ \\ \hline
\end{tabular}

\begin{tabular}{l}
\FRAME{itbpFU}{2.7769in}{2.3082in}{0in}{\Qcb{ 
\protect\begin{tabular}{l}
Contour as a function of new physics\protect \\ 
scale $\Lambda $ and $\protect\epsilon _{\protect\tau \protect\tau }^{dL}$%
\protect\end{tabular}%
}}{\Qlb{Contour with d quark}}{Figure}{\special{language "Scientific
Word";type "GRAPHIC";display "USEDEF";valid_file "T";width 2.7769in;height
2.3082in;depth 0in;original-width 3.7498in;original-height 3.7083in;cropleft
"0";croptop "1";cropright "1";cropbottom "0";tempfilename
'NFE6OX04.wmf';tempfile-properties "XPR";}}%
\end{tabular}

\begin{tabular}{l}
\FRAME{itbpFU}{2.8184in}{2.0271in}{0in}{\Qcb{ 
\protect\begin{tabular}{l}
Contour as a function of new physics\protect \\ 
scale $\Lambda $ and $\protect\epsilon _{\protect\tau \protect\tau }^{sL}$%
\protect\end{tabular}%
}}{\Qlb{s quark contour}}{Figure}{\special{language "Scientific Word";type
"GRAPHIC";display "USEDEF";valid_file "T";width 2.8184in;height
2.0271in;depth 0in;original-width 3.7498in;original-height 3.7083in;cropleft
"0";croptop "1";cropright "1";cropbottom "0";tempfilename
'NFE6KX03.wmf';tempfile-properties "XPR";}}%
\end{tabular}

It is evident from the plots above and table 1 that $\epsilon _{\alpha \beta
}^{cL}\approx \epsilon _{\alpha \beta }^{uL}\leq 10^{-2}$. As we have both
experimental and theoretical values for $K^{+}$ decay so we can specify
exact region for new physics. But for other two reactions only expected
contribution from NSIs can be given. The $D_{s}^{+}\longrightarrow D^{+}%
\overline{\upsilon }\upsilon $ and $B_{s}^{\circ }\longrightarrow B^{\circ }%
\overline{\upsilon }\upsilon $ are decays of B and charm mesons respectively
but the quark decay processes is similar to the $K$ meson decay. These are
very heavy mesons and decaying into again heavy mesons so there is a lot of
energy required for their observation. These are sensitive to $c$ quark just
like $u$ quark. We know that we have second and even third generation
constraints on free parameter of NSIs for charge leptons, like $\epsilon
_{\alpha \beta }^{e},\epsilon _{\alpha \beta }^{\mu }$and $\epsilon _{\alpha
\beta }^{\tau }$ but we have only $\epsilon _{\alpha \beta }^{uL}~$and $%
\epsilon _{\alpha \beta }^{dL}$. From the other three reactions $%
D_{s}^{+}\rightarrow K^{+}\upsilon _{\alpha }\overline{\upsilon }_{\beta
},~D^{+}\longrightarrow \pi ^{+}\nu _{\alpha }\overline{\nu }_{\beta }$ and$%
~D^{0}\longrightarrow \pi ^{0}\nu _{\alpha }\overline{\nu }_{\beta }$ we
find $\epsilon _{\alpha \beta }^{sL},$and we come to know that $\epsilon
_{\alpha \beta }^{dL}~$= $\epsilon _{\alpha \beta }^{sL}.~$So, both $~$%
generation of quarks $\left( 
\begin{array}{c}
u \\ 
d%
\end{array}%
\right) ~$and $\left( 
\begin{array}{c}
c \\ 
s%
\end{array}%
\right) $ could affect NSIs of the rare decays of mesons.

$\underset{\text{%
%TCIMACRO{\TeXButton{Table 1}{\ Table 1.}}%
%BeginExpansion
\ Table 1.%
%EndExpansion
}}{%
\begin{tabular}{|l|l|l|l|l|}
\hline
Reaction & Theoretical Br & Experimental Br & NSIs for $u$ & NSIs for $c$ \\ 
\hline
\begin{tabular}{l}
$K^{+}\longrightarrow \pi ^{+}\overline{\upsilon }\upsilon $ \\ 
$(u\overline{s})\longrightarrow (u\overline{d})\overline{\upsilon }\upsilon $%
\end{tabular}
& 
\begin{tabular}{c}
$\underset{\text{\cite{J. Brod et al.}}}{(7.8\pm 0.8)\times 10^{-11}}$%
\end{tabular}
& 
\begin{tabular}{c}
$\underset{\text{\cite{J.Beringer et al}}}{(1.7\pm 1.1)\times 10^{-10}}$%
\end{tabular}
& $2.46\times 10^{-11}$ & $2.42\times 10^{-11}$ \\ \hline
\begin{tabular}{l}
$D_{s}^{+}\longrightarrow D^{+}\overline{\upsilon }\upsilon $ \\ 
$(c\overline{s})\longrightarrow (c\overline{d})\overline{\upsilon }\upsilon $%
\end{tabular}
& $7.69\times 10^{-15}$ & not known & $2.70\times 10^{-15}$ & $2.57\times
10^{-15}$ \\ \hline
\begin{tabular}{l}
$B_{s}^{\circ }\longrightarrow B^{\circ }\overline{\upsilon }\upsilon $ \\ 
$(s\overline{b})\longrightarrow (\overline{b}d)\overline{\upsilon }\upsilon $%
\end{tabular}
& $6.86\times 10^{-17}$ & not known & $2.17\times 10^{-17}$ & $2.0\times
10^{-17}$ \\ \hline
\end{tabular}%
}$

\begin{tabular}{ll}
$\epsilon_{\tau\tau}^{uL}\sim O(10^{-2})$ & $\epsilon_{\tau\tau}^{cL}\sim
O(10^{-2})$%
\end{tabular}

$\underset{\text{%
%TCIMACRO{\TeXButton{Table 2}{\ Table 2.}}%
%BeginExpansion
\ Table 2.%
%EndExpansion
}}{%
\begin{tabular}{|l|l|l|l|}
\hline
Reaction & SM Br & NSIs for $d$ & NSIs for $s$ \\ \hline
$D^{+}\longrightarrow \pi ^{+}\nu _{\alpha }\overline{\nu }_{\beta }$ & 
\begin{tabular}{|l|l|}
\hline
Long Distance & $<8\times 10^{-16}$ \\ \hline
Short Distance & $3.9\times 10^{-16}$ \\ \hline
\end{tabular}
\cite{Slac} & $3.21$ $\times 10^{-8}$ & $3.25\times 10^{-8}$ \\ \hline
$D_{s}^{+}\longrightarrow K^{+}\nu _{\alpha }\overline{\nu }_{\beta }$ & 
\begin{tabular}{ll}
Long Distance & $<4\times 10^{-16}$\cite{TTP} \\ 
Short Distance & $1.5\times 10^{-16}$%
\end{tabular}
& $2.23\times 10^{-8}$ & $2.28\times 10^{-8}$ \\ \hline
$D^{0}\longrightarrow \pi ^{0}\nu _{\alpha }\overline{\nu }_{\beta }$ & 
\begin{tabular}{|l|l|}
\hline
Long Distance & $<6\times 10^{-16}$ \\ \hline
Short Distance & $4.9\times 10^{-16}$ \\ \hline
\end{tabular}
\cite{Slac} & $3.21\times 10^{-8}$ & $3.25\times 10^{-8}$ \\ \hline
\end{tabular}%
\ \ \ }$

\begin{tabular}{ll}
$\epsilon_{\tau\tau}^{dL}\sim1$ & $\epsilon_{\tau\tau}^{sL}\sim1$%
\end{tabular}
$~$and 
\begin{tabular}{ll}
$\epsilon_{ll^{/}}^{dL}\langle1$ & $\epsilon_{ll^{/}}^{sL}\langle1$%
\end{tabular}
for $l=l^{/}\neq\tau$

\section{Conclusion}

We have calculated NSIs Br of $B_{s}^{0}\rightarrow B^{0}\overline{\upsilon }%
\upsilon $ in NSIs and from this constraints for $\epsilon _{\tau \tau
}^{uL} $are $O(10^{-2}).~$Further, we have observed that from $%
K^{+}\longrightarrow \pi ^{+}\overline{\upsilon }\upsilon
,~D_{s}^{+}\longrightarrow D^{+}\overline{\upsilon }\upsilon $ and$%
~B_{s}^{\circ }\longrightarrow B^{\circ }\overline{\upsilon }\upsilon ~$%
constraints on $\epsilon _{\tau \tau }^{cL}~$are also $O(10^{-2})$ just like 
$\epsilon _{\tau \tau }^{uL}$. We get Br of these decays in NSIs with $c$
quark in the loop, which are exactly same as that for $u$ quark in the loop$%
. $These provide us a proof of $c$ quark induced processes for NSIs which
could affect the rare decay processes. Just like these, $D^{+}%
\longrightarrow \pi ^{+}\overline{\upsilon }\upsilon ,D^{0}\longrightarrow
\pi ^{0}\overline{\upsilon }\upsilon $ and $D_{s}^{+}\longrightarrow K^{+}%
\overline{\upsilon }\upsilon $ processes are giving $\epsilon _{\tau \tau
}^{sL}\sim 1$ and $\epsilon _{ll^{/}}^{sL}\langle 1$ exactly on equal
footing with $d$ quark. These could also affect the rare decays of charm
mesons. As a result it could safely be concluded that second generation of
quark is affecting NSIs just like first generation of quark and in NSIs
effects of second generation should be included, especially at production
level.

\end{document}